\documentstyle[preprint,aps]{revtex}
\begin{document}
\draft
\title{Aging and domain growth in the two--dimensional\\
Ising spin glass model}
\author{H. Rieger, B. Steckemetz and M. Schreckenberg}
\address{
Institut f\"ur Theoretische Physik\\
Universit\"at zu K\"oln\\
50937 K\"oln,Germany
}
\date{April 18, 1994}
\maketitle
\begin{abstract}
%
Interrupted aging in the two-dimensional Ising spin glass model 
with Gaussian couplings is established and investigated via extensive 
Monte-Carlo simulations.
The spin autocorrelation function scales with $t/\tau(t_w)$, 
where $t_w$ is the waiting time and $\tau$ is equal to $t_w$ 
for waiting times smaller than the equilibration time $\tau_{\rm eq}$.
The spatial correlations scale with $r/\xi(t_w)$,
where the correlation length $\xi$ gives information about the 
averaged domain size in the system.
Our results are better compatible with an algebraic growth
law for $\xi(t_w)$, although it can also nicely
be fitted to $(\log t_w)^{1/\psi}$ with $\psi\approx0.63$.
%
%
\end{abstract}
\pacs{75.10N, 75.50L, 75.40G.}

Aging is a characteristic feature of the non-equilibrium dynamics 
of spin glasses \cite{review}. 
After the system has been quenched into a typical
non-equilibrium state (for instance after decreasing the temperature
rapidly) the huge relaxation times at low temperatures lead to 
the experimental observation of history dependent phenomena
on laboratory time scales \cite{age_review}.
This scenario, also reported for various other strongly
disordered materials \cite{CDW_age,super_age}, is not connected to
a finite-temperature spin glass transition \cite{Schins,CuMnfilm}.
It appears that aging is just a consequence of an extremely slow 
relaxation and is either interrupted for time scales larger than the 
(possibly astronomically large) equilibration time $\tau_{\rm eq}$
or lasting forever within a spin glass phase with an infinite 
equilibration time.

It has been suggested \cite{FiHu,KoHi} that aging is a manifestation
of a slow domain growth at small temperatures, where after a
certain waiting time $t_w$ a characteristic domain size $R(t_w)$
is reached. The time dependence of subsequent observations (as for
instance the thermoremanent magnetization) show a 
clear cross-over from dynamical processes characterized by length 
scales smaller than the already achieved domain size 
(and mainly taking place inside the domains, for which reason
it is called the quasi-equilibrium regime) to processes on larger 
time scales dominated by the continuation of domain growth
through the movement of domain walls across the system.

An activated dynamics scenario \cite{FiHu} is built upon the hypothesis
that after a temperature-quench from a fully disordered state to low
temperatures domains on a length scale $L$ can grow by surmounting 
(free) energy barriers $B(L)\propto L^\psi$, where $\psi\le d-1$ is some
(temperature--independent) exponent. 
Via thermal activation such a process needs a time 
$\tau\propto\exp(\Delta B(L)/T)$. Thus after a waiting time $t_w$ the 
characteristic domain-size is given by
\begin{equation}
R(t_w)\propto(\log t_w)^{1/\psi}\;. \label{eq1}
\end{equation}
Originally such a scenario was proposed for the dynamics
of a strongly disordered system well below its phase transition 
temperature.
In the two-dimensional Ising spin glass, which does not have a finite 
temperature phase transition (see e.g.\ \cite{BhaYou}), this scenario
cannot hold for waiting times that are comparable to the 
(temperature-dependent) equilibration time $\tau_{\rm eq}$. 
However, for $t_w\ll\tau_{\rm eq}$ it might be valid
at very low temperatures. Recent experiments 
on the two-dimensional short range Ising spin glass 
Ru$_2$Cu$_{0.89}$Co$_{0.11}$F$_4$ \cite{Schins,Dekker}
indicated the validity of an activated dynamics scenario.
However, since the domain-size
or the correlation length of spatial correlations has not been
measured directly (e.g.\ via neutron scattering), 
the growth-law (\ref{eq1}) had to be verified indirectly by 
establishing the activated scaling behavior $|\log \omega| / \log t_w$
of the susceptibility $\chi_{t_w}(\omega)$ measured after aging the
system for a waiting time $t_w$. From this it was then concluded
that (\ref{eq1}) should also hold.

On the other hand, in computer simulations one has immediate 
access to the quantities of interest and the validity of a logarithmic
growth law like (\ref{eq1}) can be checked directly. Therefore we 
present in this letter the results of extensive Monte-Carlo simulations 
of the aging scenario and the domain growth in the two-dimensional 
Ising spin glass model (more details will be published elsewhere
\cite{bigone}). The latter is defined by the Hamiltonian
\begin{equation}
H=-\sum_{\langle ij\rangle} J_{ij} S_i S_j\;,
\end{equation}
where $S_i=\pm1$, the sum extends over all nearest neighbor pairs of
a $L\times L$ square lattice with periodic boundary conditions and
interaction strengths $J_{ij}$ are Gaussian random variables 
with zero mean and variance one. We choose
the usual heat bath algorithm for the spin dynamics and time is
measure in Monte-Carlo sweeps through the whole lattice.
We used lattice sizes $L$ ranging from 30 to 200 depending on 
the temperature $T$ (note that one does not need to consider
system sizes much larger than the typical correlation length to exclude
finite size effects). The quantities of interest are strongly fluctuating
from sample to sample, for which reason we averaged over several thousand
disorder realizations). The simulations were performed on a 
Parsytec GCel1024 transputer cluster.

A straightforward way to establish aging in the present model is to
calculate the spin autocorrelation function
\begin{equation}
C(t,t_w)=\frac{1}{N}\sum_i
[\langle S_i(t+t_w)S_i(t_w)\rangle]_{\rm av}\;,
\label{corr}
\end{equation}
where $\langle\cdots\rangle$ means a thermal average (i.e.\ an average
over different realizations of the thermal noise) and $[\cdots]_{\rm av}$ 
means an average over different realizations of
the bond-disorder. For each run the system is initialized in a random
initial configuration corresponding to a quench from infinite temperature
to the temperature $T$ at which the simulation is done (e.g.\ 
$\underline{S}(t_w)$ then denotes the configuration of the system 
at time $t_w$ after the quench).

In figure 1 we show the results for $C(t,t_w)$.
For higher temperatures one observes that 
curves for waiting times $t_w$ larger than a particular value, which
we denote with $\tau_{\rm eq}$, collapse. One says
that aging is interrupted here \cite{Bouch}
and this is simply a manifestation 
of the fact that the system is equilibrated now and the spin correlations
$[\langle S_i(t)S_i(t')\rangle]_{\rm av}$ become stationary in time 
(i.e.\ functions of $t-t'$ only). For lower temperatures
equilibration is not achieved on the time-scales explored
and the resulting picture is indistinguishable from that observed in
the three-dimensional Ising spin glass \cite{Rieger} or the mean 
field Ising spin glass \cite{Ritort} within the spin glass phase.
A characteristic feature of the low temperature ($T\le0.2$) behavior 
is the crossover at $t_w$ from a very slow decay at times $t<t_w$ to 
a faster, algebraic decay at times $t>t_w$:
\begin{equation}
C(t,t_w)\propto t^{-\zeta(T)}\quad{\rm for}\;t_w\ll\tau_{\rm eq}\;,
\label{corralg}
\end{equation}
which, however, is expected to
fail at much larger time-scales $t_w\gg\tau_{\rm eq}$. The exponent
$\zeta(T)$ decreases with temperature and is e.g.\ 0.05 for $T=0.2$.
Although this is a rather small exponent our data exclude a 
logarithmic decay that is predicted for activated dynamics \cite{FiHu}.

In figure 2 we show that $C(t,t_w)$ obeys simple dynamic scaling
\begin{equation}
C(t,t_w) \propto \tilde{c}[t/\tau(t_w)]\;,
\label{corrscale}
\end{equation}
where the time scale $\tau(t_w)$ is depicted in the insert. It
is equal to $t_w$ for $t_w\ll\tau_{\rm eq}$ and saturates (observable
only for high enough temperatures) at $\tau_{\rm eq}$. Note that
according to an activated dynamics scenario the scaling
variable should be $\log(t)/\log(t_w)$ rather than $t/t_w$ 
(see e.g.\ equation 3.20 in \cite{FiHu}). A $t/t_w$ scaling
has also been observed in three-dimensional spin glasses 
\cite{Rieger,expscale}, the SK-model \cite{Ritort} and simplified
spin glass models \cite{Bouch,Parisi}.

The above described aging scenario can be thought of as being  
a consequence of a very
slow domain growth, or equivalently, the slow increase of correlated
volumes in the system after the temperature quench,
as described in the introductory remarks. To verify this hypothesis
we have to look at a quantity that provides us with the necessary
informations. The domain growth in the
various strongly disordered system like the site-diluted Ising 
model \cite{DIM}, the random field Ising model \cite{RFIM} and
the random bond ferromagnetic Ising model \cite{RBFIM} or the
random bond Ising-chain \cite{Blundell} has already 
been investigated numerically. These models have the advantage that
their ground state is known to be ferromagnetic, which makes the
identification of domains easy. This is not the case for the present
system --- which is the reason why such an investigation is much 
more difficult here. One way to measure the domain size in spin glasses, 
which is however, very (computer-) time consuming, has been proposed 
in \cite{Huse} by studying the three-dimensional Ising spin glass.

We decided instead to calculate an appropriate spatial correlation 
function
\begin{equation}
G(r,t_w)=\frac{1}{t_w} \sum_{t=t_w+1}^{2t_w} 
[\langle S_0(t) S_r(t)\rangle^2]_{\rm av}\;,
\label{spatial}
\end{equation}
which is a generalization of the usual 
equilibrium correlation function
$G_{\rm eq}(r)= [ \{ {\rm Tr}_{\underline S}\,S_0 S_r 
\exp-\beta H(\underline{S})\}^2]_{\rm av}$
to the present non-equilibrium situation (note that 
$\lim_{t_w\rightarrow\infty}G(r,t_w)=G_{\rm eq}(r)$).
It is obvious that for small
waiting times $t_w$ the correlations decay rapidly to zero on the length scale
of one lattice spacing since the initial configuration is
random and spins on different sites are uncorrelated.
For increasing waiting times the system tries to relax into
energetically more favorable configurations (closer to the
--- in this case unique but unknown --- ground state). This
is the process of domain growth mentioned in the beginning
and can be read off from (\ref{spatial}) by an increase in the
number of spins that are longer correlated (to contribute to the sum
over $t$) over longer distances $r$. Thus the length scale of
the decay of $G(r,t_w)$ is a measure for the averaged domain size
$R(t_w)$.

Usually when calculating $G_{\rm eq}(r)$ the thermal average 
$\langle\cdots\rangle$ is replaced by a time average over a very 
long Monte-Carlo run. To get good statistics for the quantity
(\ref{spatial}), especially for small waiting times $t_w$, one has to 
average instead over many Monte-Carlo runs (the number of
which decreases with increasing $t_w$, fortunately) using different 
initial conditions and different thermal noise.
Furthermore one performs a spatial average by taking into account
all spin-pairs that are $r$ sites apart in the $x$-- as well as in the 
$y$--direction. To avoid
a positive bias of the statistical errors induced by the square 
(which is necessary since the site correlations alternate in sign)
we simulated two replicas $a$ and $b$ of the system simultaneously
(both with identical realization of disorder, but with uncorrelated 
initial conditions and thermal noise) and calculated the four spin 
correlations $\langle S_0^a(t) S_r^a(t) S_0^b(t) S_r^b(t)\rangle$. 
This yields results identical to those obtained by using
$\langle S_0(t) S_r(t)\rangle^2$, as we checked explicitly. 

Figure 3 shows the result for $G(r,t_w)$ for different
waiting times $t_w$ at $T=0.8$ One observes that the correlations 
grow steadily with increasing
waiting time, until, at low enough temperature, the resulting
curves collapse for $t_w>\tau_{\rm eq}$. This is in full agreement with
the picture that emerged from the time-dependent autocorrelation 
function $C(t,t_w)$ described above.
In a the linear-log plot of figure 3 the curves are only slightly
bended, meaning that the decay is roughly exponential with a 
characteristic length scale. We define an effective correlation
length (that is identical to this length scale for a pure exponential
decay) via
\begin{equation}
\xi(t_w)=2\int_0^\infty dr\;G(r,t_w)\;,
\label{xi}
\end{equation}
which, as we mentioned above, is expected to be comparable to the
averaged domain size $R(t_w)$ \cite{remark1,remark2}. 
The result is depicted in figure 4, where it can be seen that 
$\xi(t_w)$ increases slowly on a logarithmic scale until it
saturates at $\xi_{\rm eq}$ for $t_w>\tau_{\rm eq}$. 
We did a least square fit to a logarithmic growth law
\begin{equation}
\xi(t_w)-\xi_1\propto(\log t_w)^{1/\psi}\quad{\rm for}\;t_w\ll\tau_{\rm eq}
\label{logar}
\end{equation}
and obtained $\psi=0.63\pm0.05$ (roughly independent of temperature),
which is in agreement with the bound $\psi\le d-1$ and slightly 
smaller than the value found experimentally \cite{Schins}. This is
compatible with an activated dynamics scenario, see eq.\ (\ref{eq1}).
However, one obtains also a good result (in terms of the outcome
of a chi-square test) for a fit to an algebraic growth law
\begin{equation}
\xi(t_w)\propto t_w^{\alpha(T)}\quad{\rm for}\;t_w\ll\tau_{\rm eq}\,.
\label{algebr}
\end{equation}
The exponent $\alpha(T)$ decreases with temperature,
being for instance $0.08$ for $T=0.4$ and $0.04$ for $T=0.2$.
These are rather small exponents, which is the reason why it is 
hard to discriminate between (\ref{logar}) and (\ref{algebr}).

>From our point of view an algebraic domain growth yields a picture 
that is more consistent when taking into account our results for
the autocorrelation function.
On one side the proposition (\ref{eq1}) implies a logarithmic scaling of
the autocorrelation function $C(t,t_w)$, which is not in agreement what 
we find, see (\ref{corrscale}), and also implies a logarithmic decay 
$C(t,t_w)\propto \log(t)^{-\lambda'/\psi}$
for $t\gg t_w(\ll\tau_{\rm eq})$ \cite{FiHu}, which disagrees
with our result (\ref{corralg}). 
On the other hand these results (\ref{corralg}), (\ref{corrscale}) 
and (\ref{algebr}) can be explained consistently in the following way:

Let us assume a modified 
activated dynamics scenario for domain growth in which
the (free) energy barriers $B(L)$ scale with the length 
scale $L$ of the domains as $B(L)\approx a(T)\log(L)$ instead of 
$B(L)\propto L^\psi$ and let us follow \cite{FiHu} otherwise.
Then one obtains for the domain-size $R(t_w)\propto t_w^{\alpha(T)}$, 
which is exactly (\ref{algebr}) with $\alpha(T)=a(T)\Delta(T)/T$.
Furthermore from 
$C(t,t_w)\propto[R(t_w)/R(t)]^{\lambda'}$ for $t\gg t_w(\ll\tau_{\rm eq})$ 
one concludes the scaling (\ref{corrscale}) as well as the algebraic time 
decay (\ref{corralg}) with $\zeta(T)=\lambda'\alpha(T)$. 
We should mention that an algebraic growth law like (\ref{algebr})
for the domain sizes has been used by Koper and Hilhorst as
a working hypothesis in their domain theory \cite{KoHi}.

In summary we have shown that aging is present at low temperatures
even in the two-dimensional Ising spin glass model and is caused
by a very slow domain growth. We have shown that the autocorrelation 
function scales like $t/t_w$ and decays algebraically for waiting
times smaller than the equilibration time. Fitting the time dependence
of the averaged correlation length or domain size to a logarithmic
growth law or to an algebraic growth law yields reasonable results
in both cases. However, an algebraic fit is slightly better and 
is also compatible with the results for $C(t,t_w)$, whereas a
logarithmic law is not. Results for the three-dimensional 
Ising spin glass \cite{Rieger} indicate that here the situation is 
quite similar. 

We think that the simultaneous investigation
of correlations in time {\it and} in space are very important to
obtain a consistent picture of the dynamical processes in
spin glasses at low temperatures. It would be very helpful
if experimentalists could find a practicable way to gain
{\it direct} access to spatial correlation functions in spin glasses,
as is, for instance, possible in random field systems \cite{RFIMEXP}.
In numerical simulations, however,
such an endeavor is promising, as we have demonstrated here. 
For instance it would be of great interest 
to check such concepts as the overlap length in connection with 
temperature or field cycling experiments via a 
direct computation of the appropriate correlation functions in 
Monte-Carlo simulations (similar to those presented in this paper).
That might help to resolve a long lasting debate among physicists
\cite{age_review,Thermo,Granberg,Lefloch}.

We would like to thank the Center of Parallel Computing (ZPR) in K\"oln 
for the generous allocation of computing time on the transputer cluster
Parsytec--GCel1024. This work was performed within the SFB 341
K\"oln--Aachen--J\"ulich.

\begin{figure}
\caption{Autocorrelation function $C(t,t_w)$ as a function of
time $t$ for $t_w=5^n$  ($n=1,\ldots,8$) at $T=1.0$ and $0.8$, 
($n=2,\ldots,8$) at $0.6$ and $0.2$. The system size is $L=100$
and the disorder average was performed over 256 samples. The errorbars
are smaller than the symbols.}
\end{figure}

\begin{figure}
\caption{Scaling plot of the autocorrelation function $C(t,t_w)$ versus
$t/\tau(t_w)$ at $T=0.2$, The insert shows $\tau(t_w)$.}
\end{figure}

\begin{figure}
\caption{The spatial correlation function $G(r,t_w)$ for $t_w=5^n$ 
($n=1,\ldots,8$) at temperatures $T=0.8$. The insert shows 
a scaling plot of $G(t,t_w)$ versus $r/\tilde{\rho}(t_w)$. The
scaling length $\tilde{\xi}$ is equal (within the errorbars)
to $\xi(t_w)$ defined in the text and shown in the next figure.
The system size is $L=100$ and the disorder average was performed 
over 256 samples.}
\end{figure}

\begin{figure}
\caption{Log--linear plot of the correlation length $\xi(t_w)$ as 
defined in the text. The full lines are least square fits to
$(\log t_w)^{1/\psi}$ with $\psi=0.66$ for $T=0.2$, $\psi=0.60$
for $T=0.3$ and $\psi=0.63$ for $T=0.4$. The insert shows $\xi(t_w)$ 
at somewhat higher temperatures.}
\end{figure}


\begin{references}
\bibitem{remark1}
	We checked this explecitely by looking at the local 
	non-equilibrium overlaps 
	$t_w^{-1}\sum_{t=t_w+1}^{2t_w} S_i^a(t)S_i^b(t)$ as
	was done in \cite{Huse}. For details see \cite{bigone}.
\bibitem{remark2}
	The factor 2 on the r.h.s.\ in (\ref{xi}) is motivated by the fact 
	that we expect that $\int_0^\infty dr\;t_w^{-1}\sum_{t=t_w+1}^{2t_w}
	[\,|\langle S_0(t)S_r(t)\rangle|\,]_{\rm av}\approx\xi(t_w)$, 
	where the modulus takes care of the alternating
	sign of the spin correlations. Replacing the modulus by a square
	as in (\ref{spatial}) reduces the integral by a factor one half
	in case of an exponential decay.
\end{references}
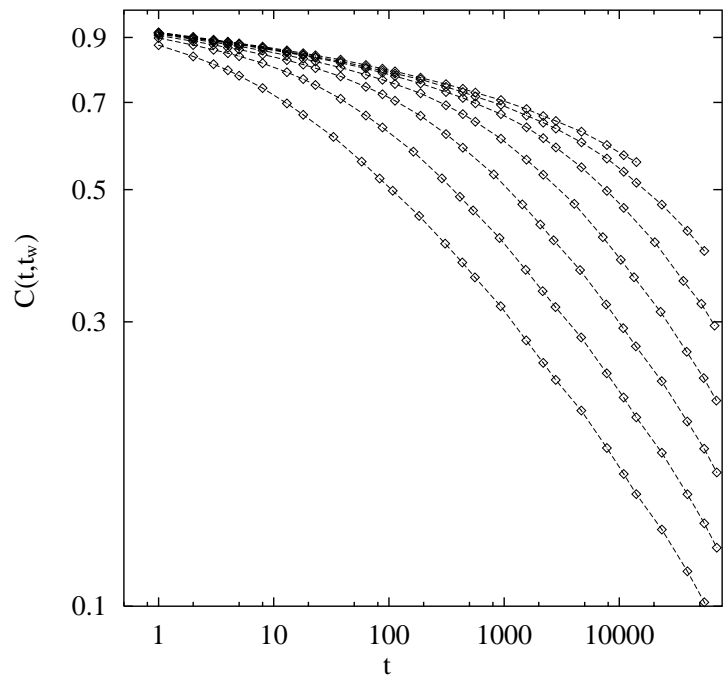
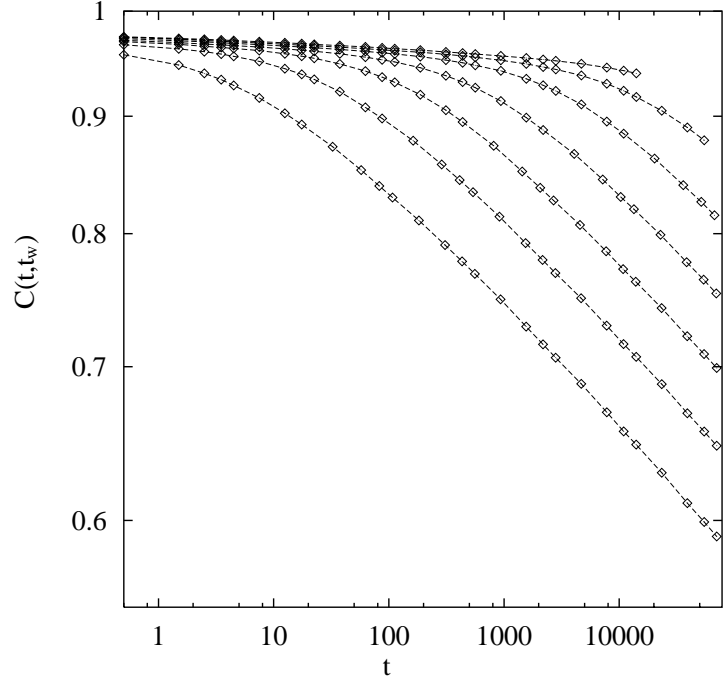
\end{document}